\pdfoutput=1
\documentclass[11pt]{article}

\usepackage[preprint]{acl}

\usepackage{amsmath}
\usepackage{times}
\usepackage{latexsym}
\usepackage{multirow}
\usepackage{booktabs}
\usepackage{url}
\usepackage{amsfonts}
\usepackage{hyperref}
\usepackage[normalem]{ulem}
\useunder{\uline}{\ul}{}
\usepackage[linesnumbered,lined,boxed,commentsnumbered]{algorithm2e} 
\usepackage{caption}
\usepackage[T1]{fontenc}
\usepackage[utf8]{inputenc}

\usepackage{microtype}

\usepackage{inconsolata}

\usepackage{graphicx}

\title{Exploring Advanced LLM Multi-Agent Systems Based on Blackboard Architecture}

\author{Bochen Han \\
  State Key Laboratory of Mathematical Sciences \\
  Academy of Mathematics and Systems Science, CAS \\
  Beijing 100190, China \\
  \texttt{hanbochen@amss.ac.cn} \\\And
  Second Author \\
  Affiliation / Address line 1 \\
  Affiliation / Address line 2 \\
  Affiliation / Address line 3 \\
  \texttt{email@domain} \\}

\author{
 \textbf{Bochen Han\textsuperscript{1}},
  \textbf{Songmao Zhang\textsuperscript{1}},
\\
\\
 \textsuperscript{1}State Key Laboratory of Mathematical Sciences \\
  Academy of Mathematics and Systems Science, CAS \\
  Beijing 100190, China \\
\\
 \small{
   \textbf{Correspondence:} \href{mailto:email@domain}{hanbochen@amss.ac.cn}
 }
}

\begin{document}
\maketitle
\begin{abstract}
In this paper, we propose to incorporate the blackboard architecture into LLM multi-agent systems (MASs) so that (1) agents with various roles can share all the information and others' messages during the whole problem-solving process, (2) agents that will take actions are selected based on the current content of the blackboard, and (3) the selection and execution round is repeated until a consensus is reached on the blackboard. We develop the first implementation of this proposal and conduct experiments on commonsense knowledge, reasoning and mathematical datasets. The results show that our system can be competitive with the SOTA static and dynamic MASs by achieving the best average performance, and at the same time manage to spend less tokens. Our proposal has the potential to enable complex and dynamic problem-solving where well-defined structures or workflows are unavailable\footnote{Code will be released soon \href{https://github.com/bc200/LbMAS/}{here}}.
\end{abstract}

\section{Introduction}

With the prevalence of large language models (LLMs), LLM-based agentic systems have attracted much attention and produced achievements in various downstream tasks including planning and reasoning \citep{putta2024agent,hao2023reasoning,sun2024thinkongraph}, code generation \citep{tang-etal-2024-codeagent,yang2024swe}, and many more. From the powerful capabilities possessed by single agent systems, LLM-based multi-agent systems (MASs) emerged, aiming to utilize collective intelligence to further enhance problem-solving performances \citep{chan2308chateval,li2023camel,li-etal-2023-theoryofmind, yin2023exchangeofthought}. 

Most LLM-based MASs proposed so far utilize fixed architectures with pre-defined agent roles and collaboration mechanisms, which often requires manual construction and thus lack generality. Some recent studies develop dynamic MASs \citep{hu2025automated,zhang2025aflow,shang2025agentsquare,liu2024dynamic,zhang2025multi}, also called autonomous MASs, which configure structures and communication strategies based on tasks and environment feedbacks. Such MASs are modularized and the optimized MAS configurations are searched in specified spaces. Compared with fixed MASs, they often have an additional, time-consuming training step and the simplified search spaces cannot cover all kinds of collaboration architectures. These two-step approaches essentially use fixed collaboration mechanisms in problem solving obtained from the supervised training based on a small number of samples in the first step.

As far back as in the 80’s of the last century, the blackboard architecture was proposed \citep{nii1986blackboard,hayes-roth_blackboard_1985} as a decentralized problem-solving approach imitating that a group of human experts work together around a shared blackboard and each contributes her/his own solutions to the blackboard until a collective decision is reached. We intend to incorporate the blackboard architecture into the LLM-based MAS and propose the blackboard-based LLM multi-agent system (bMAS) in this paper. We design a general bMAS to have three core components: (1) a control unit, (2) a blackboard, and (3) a group of LLM-based agents with different roles. For the given problem, the control unit selects the agents to participate in the current round of problem-solving according to the current messages on the blackboard. Then, each selected agent takes the contents of the whole blackboard as a prompt to its LLM and writes back to the blackboard the output produced by its LLM. Now the next round of problem-solving starts, and the process will not stop until a final solution is decided on the blackboard. 

Based on the general framework of bMAS, we present and implement a concrete bMAS in this paper and evaluate its performance on mathematical and reasoning tasks. We call it LbMAS as the first blackboard-based LLM multi-agent system. In LbMAS, the agents are designated as query-related experts and constant agents including planner, decider, critic, cleaner and conflict-resolver, and their LLMs are chosen randomly from a set of LLMs when the problem-solving process begins. The blackboard is divided into two parts, a public space and private spaces, where the former is for all agents to use and the latter solely for those involved in a debate or verification of private matter. In the blackboard system, agents communicate solely through the blackboard that contains all historical contexts. The blackboard actually serves as a public storage space to replace the memory modules typically possessed by LLM agents. The control unit of LbMAS also depends on an LLM agent to make selections for the given problem. The evaluation on various benchmarks show that LbMAS, while token-economical, obtains competitive performance when compared with both fixed and dynamic SOTA MASs.

Our contribution in this paper is two-fold. Firstly, we propose to introduce the blackboard architecture into LLM-based MAS and design bMAS as a general framework where various LLM agents communicate through a blackboard and contribute to the blackboard in an autonomous way. Secondly, we develop an implementation of bMAS called LbMAS, which, differently from other autonomous MASs, can adjust collaboration mechanisms in a timely manner according to contents of the blackboard. LbMAS has the potential to enable complex and dynamic problem-solving where well-defined structures or workflows are unavailable, just as the blackboard architecture was initially proposed for thirty years ago. And with the power of LLMs, the ancient AI architecture can revive and in turn facilitate the development of agentic systems. 

\section{Related Work}

\paragraph{Multi-Agent System.}
MAS is a computerized system composed of multiple interacting intelligent agents \citep{tran2025multi}. In this paper, we only discuss the LLM-based multi-agent systems, whose key components are introduced as follows.

• Agents: a group of LLM agents with roles, capabilities, and actions. LLM-empowered capabilities like learning, planning, reasoning and decision making  yield intelligence to the agents and overall system. Numerous agentic systems emerged for solving real-life tasks such as code generation \citep{tang-etal-2024-codeagent,yang2024swe}, medical diagnosis \citep{tu2025towards,schmidgall2025agentclinic,mcduff2025towards}, and research assistant \citep{baek-etal-2025-researchagent,schmidgall2025agentlaboratoryusingllm}.

• Environment: the external world where agents are situated and can sense and act upon. Environments can be simulated or physical spaces such as factories, roads, power grids, etc. 

• Collaboration mechanism: the pattern of collaboration and communication among agents. Generally there are two types of collaboration mechanism: cooperation \citep{islam-etal-2024mapcoder,li2023camel,shinn2023reflexion,he-etal-2023lego,10429926} and competition \citep{he-etal-2023lego,chen_llmarena_2024}.
Cooperation in an LLM-based MAS occurs when agents align their individual objectives with the same goal \citep{tran2025multi}, whereas 
competition occurs when there are conflicting objectives or scenarios with limited resources \citep{tran2025multi}. 
%In this type of interaction, the individual goals of agents clash with or oppose the objectives of others.
Debate is a specific type of competition, used when agents engage in argumentative interactions, presenting and defending their own viewpoints or solutions, and critiquing those of others, for the purpose of reaching a consensus or a more refined solution \citep{yin2023exchangeofthought,liang-etal-2024-encouraging, chan2308chateval}. 

Collaboration mechanism can also be classified from other perspectives \citep{tran2025multi}. From the structure, we can have centralized \citep{jiang-etal-2023-llmblender,qiao-etal-2024autoact}, decentralized \citep{yin2023exchangeofthought,zhang2023cumulative,chen2024agentverse,zhang-etal-2024-exploringpsychologyview}, and hierarchical collaboration mechanism \citep{chan2308chateval,li2023camel,du2023improving}. According to protocols on when agents respond and how to interact with each other, we can divide the collaboration mechanism into rule-based \citep{zhang-etal-2024-exploringpsychologyview,chen_multi-agent_2025}, role-based \citep{chen2024agentverse,hong2024metagpt,talebirad2023multi}, and model-based \citep{li-etal-2023-theoryofmind,xu-etal-2024-magic}.

The above-mentioned works use predefined collaboration mechanisms that cannot be dynamically adjusted. Recent studies focus on automatic optimization of multi-agent systems where the elements optimized are different. DsPy\citep{khattab2024dspy}  conducts prompt optimization of agents; DyLAN\citep{liu2024a}, GPTSwarm\citep{zhuge2024gptswarm} and AgentPrune\citep{zhang2025cut} focus on orchestrating interactions among agents. Specifically, DyLAN dynamically activates the composition of agents and GPTSwarm optimizes the connections between agentic nodes using a policy gradient algorithm. State-of-the-art automatic methods attempt to do more comprehensive optimizations. ADAS\citep{hu2025automated} and AFlow\citep{zhang2025aflow} achieve multi-agent workflow automation; MaAS\citep{zhang2025multi} searches for distribution of architectures rather than a single final structure. These search-based methods need a supervised training step with samples in advance and are thus not flexible enough.
\paragraph{Blackboard System.}
The Blackboard Architecture (BA) was introduced by Hayes-Roth in 1985 \citep{hayes-roth_blackboard_1985} as a complicated task-solving strategy allowing different knowledge sources to communicate by means of a common information field. It offers a compelling approach to address the control problem in expert systems \citep{muhd_mukhtar_advancing_2025}.

The BA itself relies on three core components: knowledge sources, blackboard, and control unit.
Knowledge sources are independent modules that contribute specific expertise to the problem. They do not need to know about the existence of others, but they have to understand the state of the problem-solving process and the representation of relevant information on the blackboard \citep{corkillblackboard}. Knowledge sources can be represented with different types of knowledge, including rule-based systems, case-based systems, neural networks, fuzzy logic systems, genetic algorithms, legacy software systems and others \citep{rudenko2008blackboard}.
Blackboard is used as a global database for sharing different information, such as input data, partial solutions, alternatives, and final solutions. Knowledge sources produce changes to the blackboard that incrementally lead to a solution, or a set of acceptable solutions, to the problem. The interaction among knowledge sources occurs only through changes on the blackboard \citep{nii1986blackboard}. 
%The blackboard architecture includes central blackboard architecture and distributed one\citep{jiang_constructing_2005}.
The control unit makes runtime decisions about which knowledge sources to execute next for optimal solution of the problem. \citep{corkillblackboard}. 

Blackboard systems have been designed specifically to deal with complex, ill-structured problem domains and to allow exploratory programming of knowledge-based systems by integrating heterogeneous knowledge sources \citep{cavazza2001blackboard}. This is exactly the reason we propose to introduce BA into the current LLM multi-agent systems. Note that the combination of agents with BA has been studied in \citep{1684906,article,1425173}, whereas the agents then differ fundamentally from the agents nowadays empowered by LLMs. 

%bMAS combine blackboard system with multi-agent system. It introduces the control unit to generate multi-agent invocations dynamically based on different problems, and shared blackboard space that every agent can read and write to. These modules make MAS more versatile and enable more comprehensive communication between agents. 

\section{Method}

\begin{figure*}
    \centering
    \includegraphics[width=1\textwidth]{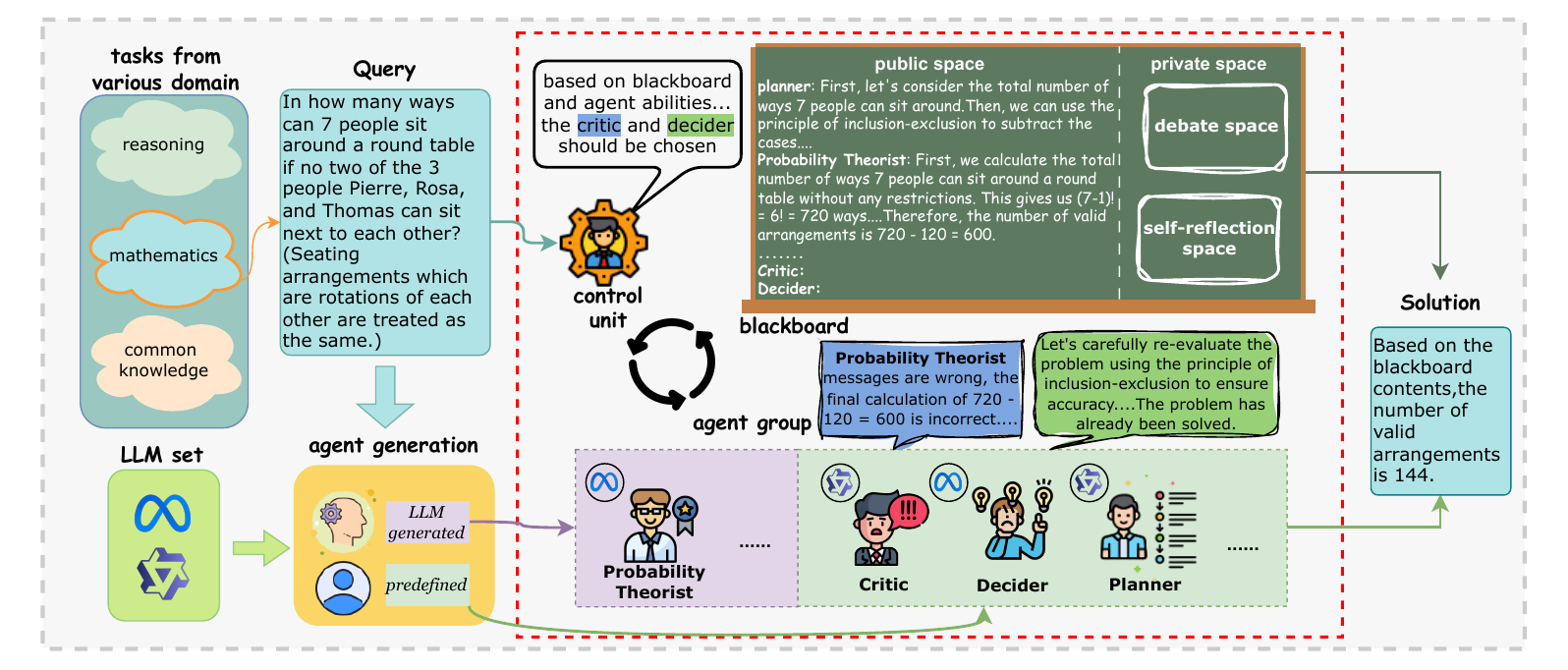}
    \caption{The general framework of bMAS is shown in the red dashed box. The illustration of LbMAS is in the green dashed box }
    \label{fig:framework}
\end{figure*}

The framework of bMAS is both illustrated in Figure~\ref{fig:framework} and presented in Algorithm ~\ref{alg:bMAS}. The blackboard is a multifunctional space including public and private parts, where the former stores dialogues and knowledge that all agents can see, and the latter allows particular agents to debate or self-reflect. The agent group contains agents with various functions, such as planning, reasoning, criticizing, tool use, etc. The control unit takes query and the current content of blackboard as input, and selects suitable agents to act in the next round of problem-solving. Such a timely, iterated process enables a dynamic adaptation of collaboration mechanism among agents. 

Further, we implement LbMAS to include an agent generation module in addition to the three general components of bMAS. The overall workflow of LbMAS is also illustrated in Figure \ref{fig:framework}. For a given query $q$, query-related agents are generated, and then the blackboard cycle starts, where the control unit iteratively selects agents from agent group to contribute to the blackboard. The cycle continues until the stopping condition is satisfied. Finally, the solution is given based on the messages on the blackboard. These steps are detailed as follows. 

\SetKwComment{Comment}{/* }{ */}
\SetKwInOut{input}{output}
\RestyleAlgo{ruled}
\begin{algorithm}
\caption{Algorithm of bMAS}\label{alg:bMAS}
    \KwIn{query $q$, agent group $G$, control unit $ConU$, maximum blackboard cycle round $K$, Blackboard contents $B$, solution extraction module $SolE$}
\KwOut{$Solution$}
$t \gets 1$\ \Comment*[r]{Current round}
\While{$t \le K$}{
    $\{A_1,...,A_j \}\gets ConU(q,B,G)$\Comment*[r]{the control unit selects agents from $G$}
    \For{$i\leftarrow 1$ \KwTo $j$}{
    $m_i \gets Exec(A_i,B)$ \Comment*[r]{execute agent to generate message}
    $ B \gets B\cup m_i$\Comment*[r]{update blackboard contents}
    }
    $t \gets t+1$
    }
$Solution \gets SolE(B, G)$\Comment*[r]{get $solution$ from the blackboard}
\end{algorithm}

\subsection{Generating agents}
When dealing with diverse queries from different domains, it is advisable to generate query-related expert identities and instruct LLMs to respond as the generated experts \citep{long-etal-2024-multiexpertprompting,xu2025expertpromptinginstructinglargelanguage}. Following this, we conduct the agent generation step in LbMAS. When generating an LLM agent there are two main factors that can be customized: the prompt and base LLM. We only consider role prompts and an agent generating agent (AG) is executed to randomly generate $n$ expert prompts from domains related to query $q$ based on the expert generation instruction $I$.
\begin{equation}
\{(E_1, D_1),...,(E_n, D_n) \} = AG(q, I) \label{ag}   
\end{equation}
%%$$ \{(E_1, D_1),...,(E_i, D_i) \} = AG(q, I)$$
Each expert prompt is a tuple $(E_i,D_i)$ where $E_i$ is the expert’s identity and $D_i$ a short description of its expertise, by which expert agents are then generated. For the base model, the study \citep{zhang_if_2025} shows that MASs using different LLMs rather than a single one have better performance since model diversity can effectively compensate for individual model limitations while amplifying their strengths. We thus create a set of LLMs, and every time a new agent is created, it will randomly choose an LLM from the set as its base model.
\subsection{The blackboard cycle}
There are three modules involved in the blackboard cycle and we present them subsequently.
\paragraph{Agent group.}
The agent group in LbMAS contains a set of predefined agents in addition to query-specified agents. 
%the former generated in the agent generation step whereas the latter ensuring the autonomous and effective operation of the system. 
In the original blackboard systems messages are generated and transferred following rules and logical inferences, which in LbMAS can be realized by the predefined LLM agents that are decider, planner, critic, conflict-resolver, and cleaner. The decider agent assesses whether messages on the blackboard are enough to yield the final solution; and if so, it will give the solution and the blackboard cycle stops. The planner agent makes plans to solve the query, and particularly would decompose the query into smaller tasks when it is complex. The critic agent points out errors in messages on the blackboard, which may come from hallucinations of LLMs, and forces relevant agents to rethink their output. The conflict-resolver agent detects contradictions among messages on the blackboard and then name the agents that have conflicting messages. These called agents would discuss about the conflicts, which can be conducted in the private space of the blackboard. After discussion, each involved agent generates a new message and writes it in the public blackboard space. The cleaner agent is typically designed to ensure the quality of messages on the blackboard. It detects useless or redundant messages and removes them. The cleaning is necessary as effective management of content of the blackboard can facilitate meaningful communication among agents and at the same time reduce token consumption. 

%These agents are selected based on previous MAS and the need for LbMAS. Once agents selected by the control unit, it will generate response according to messages on the blackboard, formally:
%\begin{equation}
%m = A(q, M) \label{aa}
%\end{equation}
%%$$m = A(q, B)$$

%where q is the input query and B is the contents on the blackboard public space.

\paragraph{Blackboard.}
The blackboard public space serves as a shared memory in which each LLM agent can read and write, enabling seamless communication and collaboration. Agents can access and utilize messages stored in the public space, allowing them to incrementally build upon each other's results. Unlike MetaGPT \citep{hong2024metagpt} that uses a shared memory pool as an assistant for knowledge storage, in LbMAS agents communicate solely through the blackboard without any direct contact; in other words the blackboard is responsible for all agent communication and agents decide on their own what to write on the blackboard.

With the availability of blackboard, we remove the memory module commonly existing in LLM-based agents. The main function of the memory module is to ensure that the agent remembers the context and makes consistent output. In LbMAS agents' messages are all stored on the blackboard and the memory modules become unnecessary. This way agents can ensure the consistency of each message while the overall length of prompts for all agents can be reduced, enabling a MAS to have more discussions under the token constraint.

%Mostly agents only access the public space. In LbMAS the private space is only used when agents selected by the conflict resolver need discussion. After 
\paragraph{Control unit.}
The control unit $ConU$ manages the flow of information and the execution of tasks among agents, pursuing that right actions are taken according to current messages on the blackboard. Again we use an LLM to fulfill these functions. When the cycle starts, the control unit iteratively selects agents based on query $q$, current messages $B$ on the blackboard and agent abilities $D_1,...,D_n$.
\begin{equation}
\{E_{i_1},E_{i_2},...,E_{i_j} \} = ConU(q, B, \{D_1,...,D_n \})\label{cc}
\end{equation}
%% $$ \{A_1,A_2,...,A_j \} = C(q, B, \{D_1,...,D_n \})$$
 %Where $B$ is the current blackboard contents, $\{D_1, ...,D_n \}$ are short descriptions of the ability of the agents, including experts and other agents. The chosen agents then generate information and individually add it to the blackboard. 
When the maximum iteration number is reached or the decider agent is selected and gives the solution, the cycle comes to an end. Normally there are two ways for a MAS to yield the final answer, depending on a specific agent to make decisions, or conducting a majority vote in which all relevant agents can participate. We have implemented both: the decider will give the answer when possible, and when the cycle stops all agents are asked to present an answer based on current content of the blackboard. The cumulative similarity is calculated for each answer relative to the others, denoted as $V\left(a_{i}\right)=\sum_{j=1, j \neq i}^{N} \operatorname{sim}\left(a_{i}, a_{j}\right)$ where $N$ is the number of answers obtained. The one with the highest cumulative similarity is then chosen as the final answer, i.e.,  $argmax_{a_i}V(a_i)$. 

%When the blackboard cycle stops, the system should get the final answer based on the blackboard contents. Let $A$ represent the final answer. There are two commonly used methods: (1) use a specified agent to output an answer. (2) Conduct a majority vote in which all agents can participate. We have implemented both methods. The decider will give the answer when possible. What's more, when the blackboard cycle stops we ask all agents to give an answer based on blackboard contents. The cumulative similarity is calculated for each answer relative to the other answers, denoted as $V\left(a_{i}\right)=\sum_{j=1, j \neq i}^{N} \operatorname{sim}\left(a_{i}, a_{j}\right)$. The answer with the highest cumulative similarity is then chosen as the final answer denoted as $A=argmax_{a_i}V(a_i) $. 
\section{Experiments}

\begin{table*}[]
\caption{Comparing LbMAS with single-agent and static multi-agent systems. The best result are highlighted in bold and the runners-up are underlined. }
\centering
\scalebox{0.9}{
\begin{tabular}{@{}l|cccccc|c@{}}
\toprule
\textbf{Method}     & \textbf{MMLU}  & \textbf{ARC-Challenge} & \textbf{GPQA-Diamond} & \textbf{BBH} & \textbf{MATH} & \textbf{GSM8K} & \textbf{Avg.}  \\ \midrule
Vanilla (Llama)           & 78.16          & 90.87                  & 36.86                 & 64.00        & 26.00            & 34.57          & 54.74          \\
Vanilla (Qwen)            & 76.40          & 89.50                  & 36.36                 & 74.40        & 40.20          & 42.83          & 59.94          \\
CoT (Llama)          & 81.31          & 92.74                  & 35.35                 & 81.20        & 62.20          & \textbf{94.84} & 74.60          \\
CoT (Qwen)           & {84.82}    & 92.74                  & 44.94                 & {\ul 87.20}        & \textbf{76.40} & 94.46          & 80.09          \\ \midrule
Major-Vote          & 81.49          & 91.97                  & 41.41                 & 73.20        & 36.00            & 39.80          & 60.64          \\
MultiPersona        & 81.84          & 90.35                  & 44.44                 & 85.60        & 65.80          & 93.85          & 76.98          \\
Exchange-of-Thought & 81.05          & 91.12                  & 37.87                 & 80.80        & 60.40          & 83.32          & 72.42          \\
Chateval            & 84.21          & {93.25}            & {\ul 51.26}                 & \textbf{90.00}  & 70.20          & 94.46          & 80.56          \\ \midrule
\textit{LbMAS}                 & \textbf{85.35}          & {\ul 93.43}                  & \textbf{54.04}           & \textbf{90.00}  & {\ul 72.8}            & 94.46          & \textbf{81.68}    \\
\textit{LbMAS-Majorvote}           & {\ul 85.17} & \textbf{93.51}         & \textbf{54.04}        & \textbf{90.00}  & {72.60}    & {\ul 94.61}    & {\ul 81.65} \\ \bottomrule
\end{tabular}
}
\label{tab:main}
\end{table*}

\begin{table}[]
\caption{Comparing systems with a single base LLM in mathematical benchmarks. The best results are highlighted in bold.}
\centering
\begin{tabular}{@{}lcc@{}}
\toprule
\textbf{Method} & \textbf{MATH} & \textbf{GSM8K} \\ \midrule
Vanilla (Llama)  & 26.0            & 34.57          \\
CoT (Llama)      & 62.2          & 94.84          \\
\textit{LbMAS (Llama)}    & 58.4          & 92.57          \\ \midrule
Vanilla (Qwen)   & 40.2          & 42.83          \\
CoT (Qwen)       & 76.4          & 94.46          \\
\textit{LbMAS (Qwen)}     & \textbf{78.6} & \textbf{96.05} \\ \bottomrule
\end{tabular}
\label{tab:single}
\end{table}

\subsection{Implementation Setup}

\paragraph{Tasks and benchmarks.}
We evaluate LbMAS on six benchmarks: the knowledge dataset MMLU \citep{hendrycks2021measuring}, reasoning datasets ARC-Challenge \citep{clark2018thinksolvedquestionanswering}, GPQA-Diamond \citep{rein2024gpqa} and BBH-dateunderstand \citep{suzgun-etal-2023-challengingbbh}, and mathematical datasets MATH \citep{hendrycks2021measuringmath} and GSM8K \citep{cobbe2021trainingverifierssolvemath}. For MMLU, we randomly select 20 questions from each category and obtain a total of 1140 questions. For MATH, we use a subset of 500 questions randomly selected by OpenAI\footnote{https://huggingface.co/datasets/HuggingFaceH4/MATH-500}. 
%The dataset statistics are in Table~\ref{tab:dataset}. 
\paragraph{Baselines.}
We compare our method with Chain-of-Thought (CoT) \citep{wei2022chain}, Major-Vote \citep{li2024more},  static multi-agent systems including Chateval \citep{chan2308chateval}, Exchange-of-Thought (EoT) \citep{yin2023exchangeofthought}, and MultiPersona (MP) \citep{liang-etal-2024-encouraging}, and autonomous multi-agent systems including GPTSwarm \citep{zhuge2024gptswarm}, AFlow \citep{zhang2025aflow}, and MaAS \citep{zhang2025multi}. 
%More details of baseline setups are provided in Appendix.
\paragraph{Implementation details.}
Our main experimental results were obtained using two open-source LLMs with varying sizes and capacities: Llama-3.1-70b-Instruct \citep{DBLP:journals/corr/abs-2407-21783} and Qwen-2.5-72b-Instruct \citep{qwen2025qwen25technicalreport}. To eliminate the impact of model differences, when creating a new agent we randomly select one of the two LLMs as its base model. For the sake of fairness, all MASs in comparison have adopted this LLM selection way. All models are accessed via APIs with the temperature set to 0.7. We set the maximum rounds of the blackboard cycle to 4. 
%We also conduct experiments with smaller and larger LLMs. Results are in Appendix.

\subsection{Performance Analysis}

The main results of the experiments are shown in Table~\ref{tab:main}. 
%The results demonstrate the effectiveness of LbMAS. 
One can see that LbMAS obtained the best performance on most test benchmarks, outperforming CoT methods by an average of 4.33\% and static methods by 5.02\%. The two bottom lines of Table~\ref{tab:main} compare the performance of two decision-making ways LbMAS adopts. The decider and majority vote way performed similarly on all benchmarks, indicating that in most cases the decider and other agents reached a consensus after discussion. As a matter of fact, the proportions of consensus reached by all agents on the MMLU, GPQA and MATH datasets are 89.8\%, 52.5\% and 29.4\%, respectively. 

Note that in mathematical problems, the CoT method using Qwen LLM achieved the best performance. To further analyze, we conducted a comparison using one base model for all agents as shown in Table~\ref{tab:single}. Our system solely using Qwen outperformed the CoT method on the MATH and GSM8K benchmarks, whereas LbMAS with Llama as the only base model performed worse than CoT. The results indicate that the Llama model may not be as good as Qwen at handling mathematical problems that usually require multi-step reasoning and accurate calculations, and such a performance gap among LLMs can affect the overall performance of MASs.

%We compare the performance of two decision-making methods on six benchmarks. results show that in bMAS major-vote and one decision-making agent have similar performance, indicating that in most cases the decider and most other agents reached a consensus after discussion. Figure~\ref{fig:consensus} further confirms our conclusion. The proportions of consensus reached by all agents on the MMLU, GPQA and MATH datasets are 89.8\%, 52.5\% and 29.4\%, respectively. 

\begin{figure}[htbp]
    \centering
    \includegraphics[width=1\linewidth]{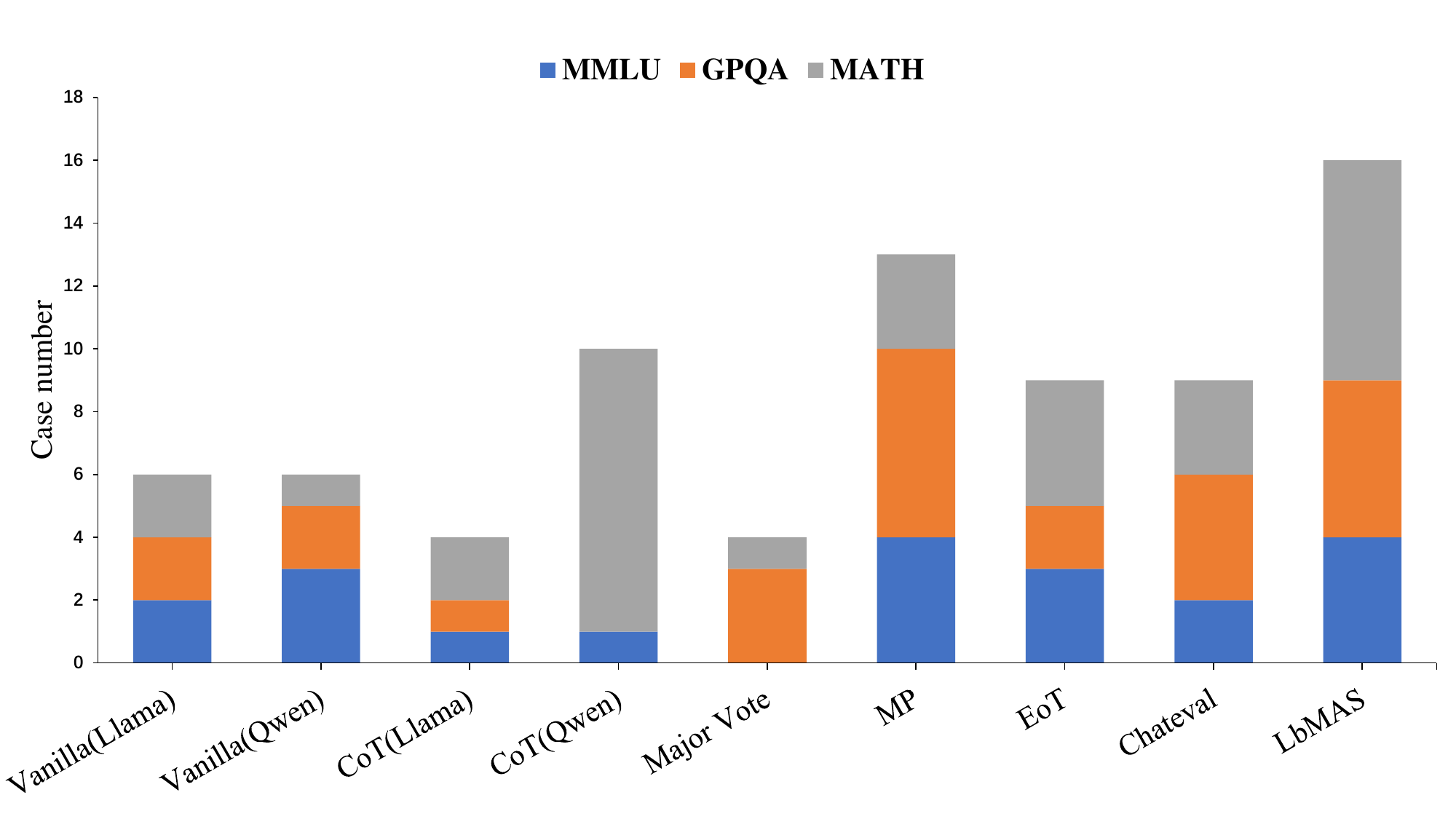}
    \caption{The number of queries answered correctly by a single method}
    \label{fig:unique}
\end{figure}

We also counted the number of queries answered correctly by a single method, as shown in Figure~\ref{fig:unique}. Our method uniquely solved the highest number of problems overall in the MMLU, GPQA and MATH datasets, while other methods including Vanilla are all able to uniquely solve problems, showing the respective potential of each method. 

\begin{table*}[]
\caption{Comparing the token cost and performance of bMAS, static MAS and autonomous MAS on the MATH dataset. The lowest total tokens and highest performances are highlighted in bold and the second ones underlined.}
\centering
\resizebox{\textwidth}{!}{%
\begin{tabular}{@{}lcccccccc@{}}
\toprule
\textbf{Method} & \multicolumn{2}{c|}{\textbf{Training}} & \multicolumn{2}{c|}{\textbf{Inference}} & \multicolumn{4}{c}{\textbf{Overall}} \\ \cmidrule(l){2-9} 
 & \begin{tabular}[c]{@{}c@{}}Prompt\\ token\end{tabular} & \multicolumn{1}{c|}{\begin{tabular}[c]{@{}c@{}}Completion\\ token\end{tabular}} & \begin{tabular}[c]{@{}c@{}}Prompt\\ token\end{tabular} & \multicolumn{1}{c|}{\begin{tabular}[c]{@{}c@{}}Completion\\ token\end{tabular}} & Total prompt & Total completion & Total token & Performance \\ \midrule
EoT & - & - & 5,475,154 & 2,853,618 & 5,475,154 & 2,853,618 & 8,328,772 & 60.40 \\
MP & - & - & 1,351,821 & 820,069 & {\ul 1,351,821} & \textbf{820,069} & \textbf{2,171,890} & 65.80 \\
Chateval & - & - & 3,338,508 & 2,113,312 & 3,338,508 & 2,113,312 & 5,451,820 & 70.20 \\ \midrule
GPTSwarm & 445,674 & 1,961,010 & 559,728 & 2,483,013 & \textbf{1,005,402} & 4,444,023 & 5,449,425 & 67.25 \\
AFlow & 6,643,444 & 7,977,945 & 1,562,388 & 513,255 & 8,205,832 & 8,491,200 & 16,697,032 & 69.25 \\
MaAS & 4,150,172 & 2,700,925 & 3,773,460 & 2,388,101 & 7,923,632 & 5,089,026 & 13,012,658 & {\ul 70.77} \\ \midrule
\textit{LbMAS} & - & - & 3,352,594 & 1,013,015 & 3,672,222 & {\ul 1,049,267} & {\ul 4,721,489} & \textbf{72.60} \\ \bottomrule
\end{tabular}%
}
  \label{tab:cost}
\end{table*}

\subsection{Cost Analysis}

We compared the token cost as well as the performance between LbMAS and the static and autonomous MAS baselines on the MATH benchmark, as shown in Table~\ref{tab:cost}. One can see that LbMAS not only stood out in performance but also managed to cost the second lowest number of tokens. 
%We observe that LbMAS is efficient and high-performance, outperforming both autonomous and static methods on MATH dataset. 
Note that the three autonomous MASs take a two-step strategy to search for an optimized workflow first before solving the problem, thus spend a lot more extra tokens. In addition, they require training data from related domains to perform searching and shall become useless when such data is unavailable. 
Our method on the other hand takes a dynamic adjusting strategy to select and execute agents on-the-fly according to the current blackboard messages in an iterative way until a consensus solution is obtained.
%LbMAS integrates them into one step, dynamically adjusts agent invocation based on current blackboard messages while executing, which saves token consumed in the search step while having competitive performance. In addition, two-step autonomous MAS require training data from related domains to perform searching and can not handle the situation where only the test set is available.

\subsection{Case Study}

\begin{figure*}[htbp]
    \centering
    \begin{minipage}[t]{0.9\textwidth} 
      \centering
      \includegraphics[width=0.9\textwidth]{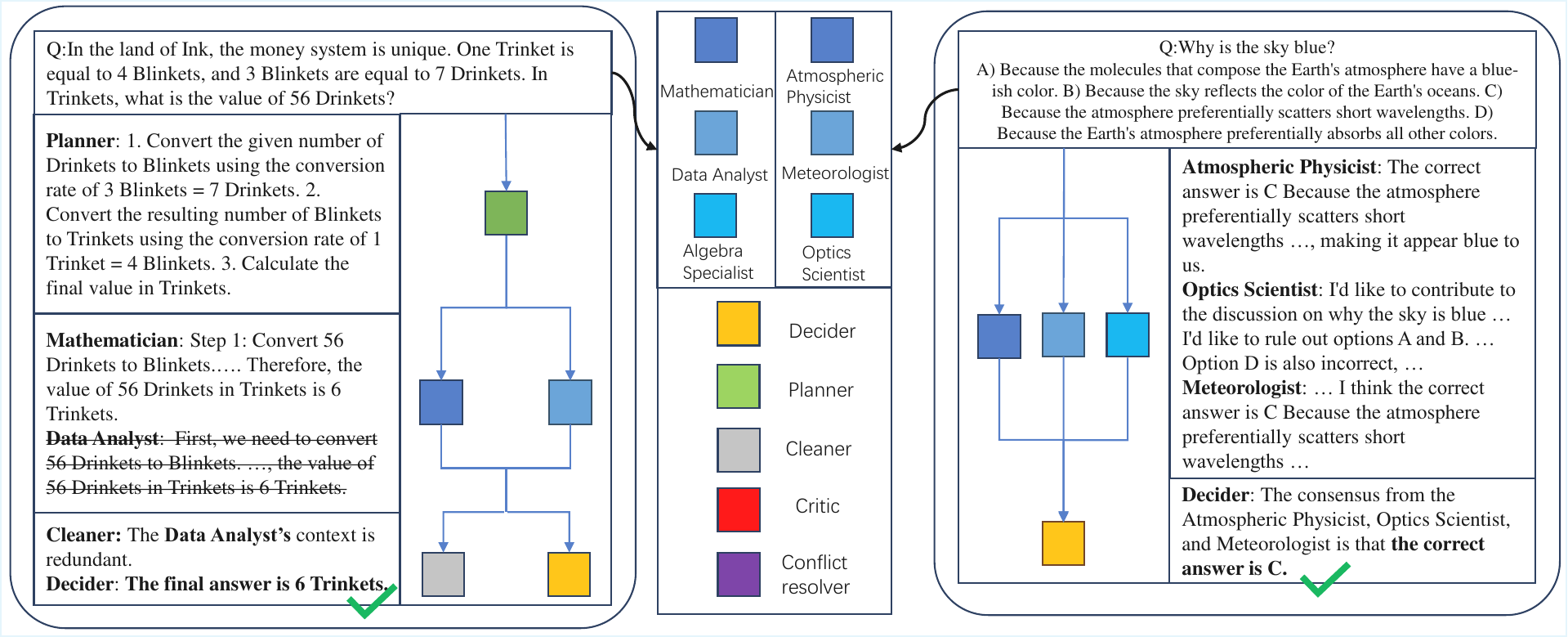}
      \caption*{Easy cases a) on the left and b) on the right}
  \end{minipage}
  \begin{minipage}[t]{0.9\textwidth}
      \centering
      \includegraphics[width=0.9\textwidth]{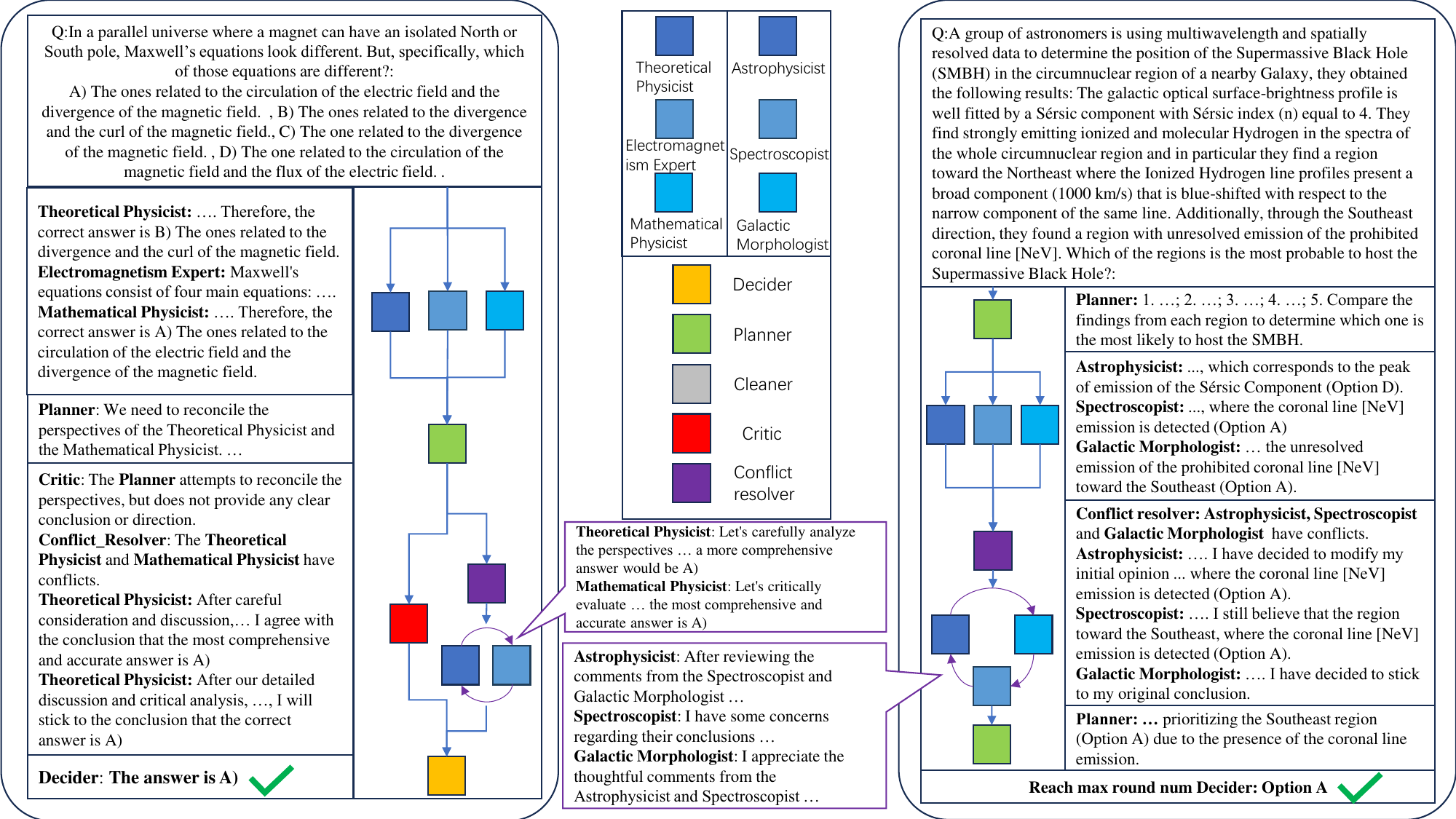}
      \caption*{Hard cases c) on the left and d) on the right}
  \end{minipage}
  \caption{Case study of LbMAS.}
    \label{fig:case}
\end{figure*}

 We visualize the process of using LbMAS to solve four problems of different difficulty levels in Figure~\ref{fig:case}. When facing easy problems, LbMAS can stop after one round of execution, while for hard problems LbMAS would have agents engage in multiple rounds of discussion to obtain the final answer. This shows that our method can generate suitable multi-agent collaborations for diverse problems. In addition, agents can effectively play their roles, as shown in query a) where the cleaner detects and removes redundant messages and in query c) where the conflict resolver identifies agents with conflicting messages and asks them to have a discussion.

\subsection{Framework Study}

\paragraph{Round analysis.}
We analyze the impact of the maximum number of blackboard cycle rounds on the system performance, as illustrated in Table~\ref{tab:round}. For the three datasets MMLU, GPQA and MATH, the average rounds are 2.88, 3.05, and 3.29 respectively. This is why we set the maximum round number to 4. Additionally, we find that easy problems such as those in MMLU need lower numbers of rounds whereas hard problems such as GPQA and MATH require more rounds of execution.
%the average number of rounds reaches a bottleneck as the setting increases, and this bottleneck is related to the dataset. MMLU where the problems are simple has a low average number of rounds, while GPQA and MATH with relatively hard problems have a larger average number of rounds. The results also demonstrate that a larger number of rounds on the MATH dataset can improve performance.

\paragraph{Ablation study.}
We performed an ablation study on the control unit to test the autonomy of LLM agents in LbMAS. The agents are instructed to determine on their own whether to respond to the current content on the blackboard without the control unit. The results in Table~\ref{tab:ablation} show that the control unit can significantly reduce the token cost of LbMAS with a slight turbulence in performance. Competitive results indicate that LLM agents have a certain degree of autonomy and that the control unit can select appropriate agents in each round. Further, the bottom line in Table~\ref{tab:ablation} shows that the performance declines on three datasets if the cleaner agent is asked to mark the redundant messages rather than directly remove them, indicating the necessity of managing the content of blackboard.

\begin{table}[]
 \caption{Comparing the average execution rounds of the blackboard cycle and performances (in parentheses) under different maximum round settings. The best performances are highlighted in bold.}
  \centering
\scalebox{0.85}{
\begin{tabular}{@{}cccc@{}}
\toprule
\textbf{Max Round} & \textbf{MMLU} & \textbf{GPQA} & \textbf{MATH} \\ \midrule
2                      & 2.00 (84.28)    & 2.00 (52.02)   & 2.00 (72.80)   \\
4                      & 2.88 (\textbf{85.17})   & 3.29 (\textbf{54.04})   & 3.04 (72.60)   \\
6                      & 3.06 (84.21)   & 3.44 (53.03)   & 3.34 (71.60)   \\
8                      & 3.15 (84.21)   & 3.46 (51.01)   & 3.35 (\textbf{74.80})   \\ \bottomrule
\end{tabular}
}
  \label{tab:round}
\end{table}

\begin{table}
 \caption{Ablation study of LbMAS. The best performances are highlighted in bold.}
  \centering
\scalebox{0.65}{
\begin{tabular}{@{}llccc@{}}
\toprule
\textbf{Method}                                                              & \textbf{}   & \textbf{MMLU}  & \textbf{GPQA}  & \textbf{MATH}  \\ \midrule
\multirow{2}{*}{LbMAS(full)}                                                 & performance & 85.17          & \textbf{54.04} & 72.60          \\
                                                                             & token cost  & 5,074,931      & 2,984,204      & 4,721,489      \\ \midrule
\multirow{2}{*}{\begin{tabular}[c]{@{}l@{}}w/o control \\ unit\end{tabular}} & performance & \textbf{85.52} & 53.53          & \textbf{72.80} \\
                                                                             & token cost  & 18,829,644     & 13,325,826     & 13,858,631     \\ \midrule
\begin{tabular}[c]{@{}l@{}}w/o message \\ removal\end{tabular}               & performance & 84.38          & 53.53          & 71.00          \\ \bottomrule
\end{tabular}
}
  \label{tab:ablation}
\end{table}

\renewcommand{\dblfloatpagefraction}{.9}

\section{Discussion}

\paragraph{Advancement of the blackboard-based LLM multi-agent framework.}
We summary the advancement of LbMAS as follows. Firstly, our framework is capable of dynamically selecting agents, making it a highly generalized framework that can solve multi-domain problems without the need for additional prompt modifications or collaboration mechanism optimization. Secondly, our framework enables a high degree of autonomy in both agent generation and interaction, and agents with various roles can complete their tasks properly. %Regardless of the size of the base LLM, small LLMs with relatively poor capabilities can also perform well. 
Thirdly, our framework has high scalability as there are no constraints about the agent group, making it easy to add new agents and functions whenever they are needed. Meanwhile, as a medium for agent interaction, the blackboard can meet various kinds of interaction needs including public vs. private space, flat vs. hierarchical structure, and so on. 

\paragraph{Why does the blackboard-based LLM multi-agent system work?}
Compared with the SOTA static and autonomous MAS methods, we analyze that the performance improvement and cost saving of LbMAS can come from two aspects: (1) a shared memory pool by the blackboard, which enables more comprehensive information exchange among agents, and (2) dynamically selecting agents from the agent group based on the current blackboard, which can ensure that the selected agents are suitable for handling existing messages and thus avoid unnecessary agent executions. Particularly, unlike recent autonomous MASs that search for optimal workflows first before running them to solve problems, our system adapts the agent selection and execution on-the-fly alongside the changing on the blackboard. Although in the end the whole execution order may not be optimal, we do not need a supervised training step that requires extra data and token cost. 

\paragraph{Conclusions.}
In this paper, we propose the blackboard-based LLM multi-agent framework, which integrates flexibility and autonomy of the traditional blackboard architecture into nowaday MASs, aiming to enable dynamic and efficient problem-solving when workflows are unknown. We develop the first implementation and conduct experiments on six knowledge, reasoning and mathematical datasets. The results have demonstrated the cost-effectiveness of the system and we believe that our proposal shall contribute to the development of fully automated and self-organizing MASs.

\section*{Limitations}

Our work is the first attempt to combine blackboard architecture with LLM multi-agent systems, and there are limitations in the implementation. Firstly, there are relatively fewer predefined types of agents in the agent group. We implemented agents with roles, and commonly used functions such as code writing and tool usage are not yet considered. In addition, experiment benchmarks are limited, and conducting experiments on a wider range of datasets can explore the potential of the framework, which will be one of our future works. 
Secondly, the agent generation module is relatively simple without considering enough factors. Results on the mathematical datasets show the necessity of selecting suitable LLMs, and works like MASRouter \citep{yue2025masrouterlearningroutellms} that can optimize agent prompts and base LLMs in the agent generation process should be applied to our system.  
Thirdly, exploration of the blackboard is not adequate. Comparative experiments should be conducted to explore the differences between shared memory pool and agent memory module, as memory of LLM agent is the key component to support agent-environment interactions \citep{zhang2024surveymemorymechanismlarge}.

\bibliography{anthology,custom}
%\bibliography{custom}

\appendix

\section{Experiment details}
\subsection{Datasets}
In Table~\ref{tab:dataset}, we provide an introduction to the datasets used in our experiments, including problem type, answer format, and test samples. Note that in MATH dataset the answer format is varied, including number, word and expression. Accuracy is used as the evaluation metric in our experiments. We require the intelligent agent to output answers in "boxed{answer}" format for the convenience of extracting answers. For datasets with multiple-choice answer format, accuracy is calculated by checking if the option extracted from the response matches the correct answer. For other format we perform a match with the ground-truth answer.

\subsection{Baselines}
In this section, we provide a description of the configurations for baseline MAS methods. For static MAS MultiPersona, Exchange-of-Thought, and Chateval, we utilize the implementation from\citep{zhang_if_2025}. Three autonomous MAS methods are implemented in accordance with the original implementation\citep{zhang2025aflow, zhuge2024gptswarm,zhang2025multi}.

\subsection{Prompts}
In this section, we present the prompts for each module. The Agent generation prompt is based on Multi-expert Prompting\citep{long-etal-2024-multiexpertprompting}. The system prompt is unique to every predefined agent and includes and the query and the agent's role. Generated experts use the roles generated by the agent generation module.

\begin{figure*}[htbp]
    \centering
    \includegraphics[width=0.9\linewidth]{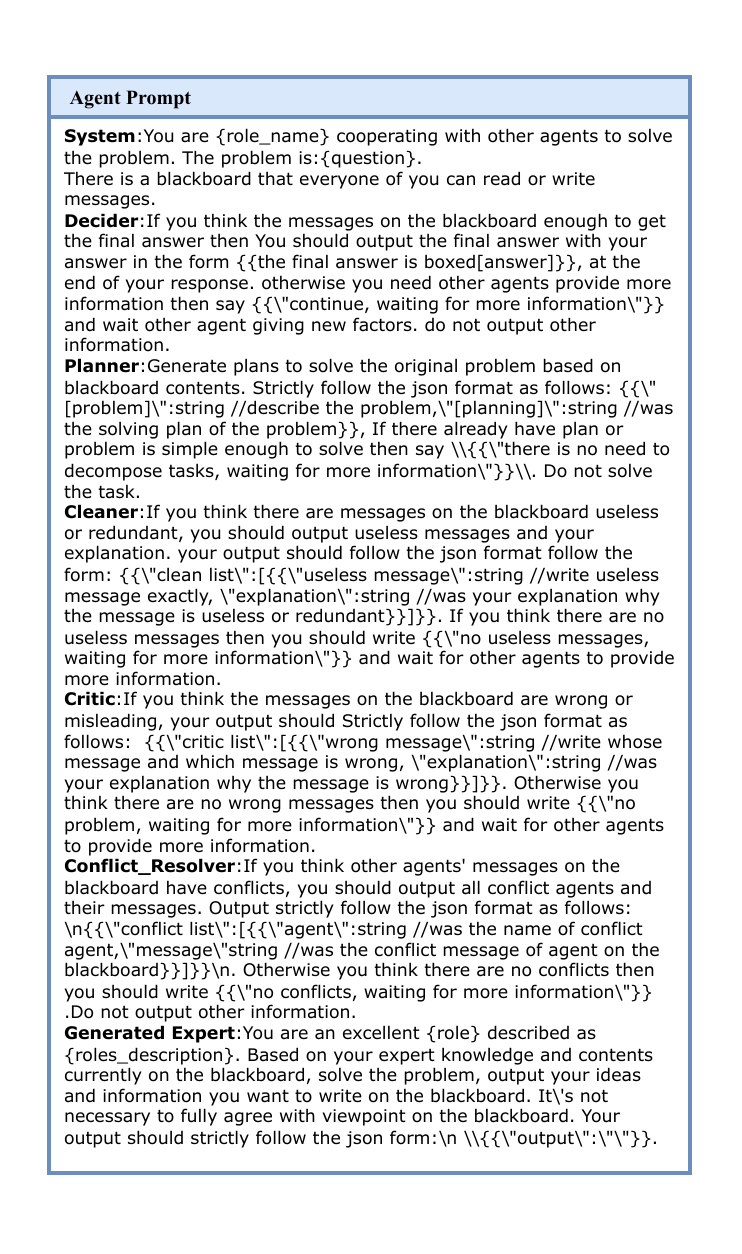}
    \caption{Agent prompt}
    \label{fig:agent}
\end{figure*}

\begin{figure*}[htbp]
    \centering
    \includegraphics[width=1\linewidth]{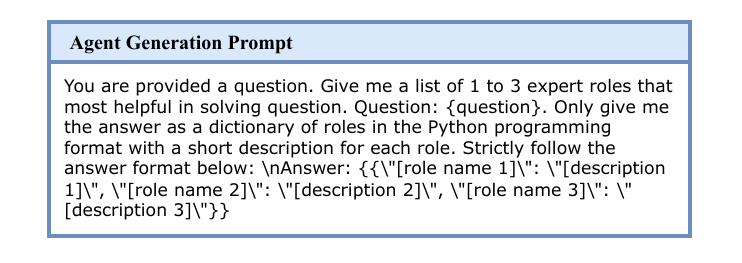}
    \caption{Agent-generation prompt}
    \label{fig:ag}
\end{figure*}

\begin{figure*}[htbp]
    \centering
    \includegraphics[width=1\linewidth]{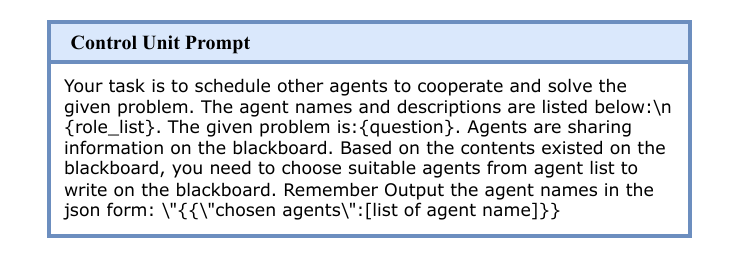}
    \caption{Control unit prompt}
    \label{fig:control}
\end{figure*}

\begin{table*}[]
 \caption{statistics of datasets utilized in our experiment.} \centering
\begin{tabular}{@{}llcl@{}}
\toprule
DATASET            & TYPE                   & ANSWER FORMAT & TEST \\ \midrule
MMLU               & General Knowledge      & Multi-choice  & 1140 \\
ARC-Challenge      & Scientific Reasoning   & Multi-choice  & 1172 \\
GPQA-Diamond       & Scientific Reasoning   & Multi-choice  & 198  \\
Date-Understanding & Symbolic Reasoning     & Multi-choice  & 250  \\
MATH               & Mathematical Reasoning & Uncertain     & 500  \\
GSM8K              & Mathematical Reasoning & Number        & 1319 \\ \bottomrule
\end{tabular}
 \label{tab:dataset}
\end{table*}

\section{Additional experiments}

Table~\ref{tab:totalsingle} shows the performance comparison of LbMAS based on a single LLM and two LLMs. Generally, LbMAS with two LLMs outperforms the average of a single LLM. In GPQA-Diamond and MMLU datasets LbMAS has the best performance. Although experimental results have proven the generalization LbMAS, we can see that in relatively easy dataset MMLU, the performance improvement brought about by the LbMAS is limited. It is advisable to test LbMAS on more types of tasks.

\begin{table*}[]
\caption{Comparing systems with a single base LLM.}
\centering
\begin{tabular}{@{}lcccc@{}}
\toprule
Method             & MMLU  & ARC-Challenge & GPQA-Diamond & BBH   \\ \midrule
CoT (Llama)        & 81.31 & 92.74         & 35.35        & 81.20 \\
CoT (Qwen)         & 84.82 & 92.74         & 44.94        & 87.20 \\ \midrule
LbMAS (Llama)      & 81.22 & 91.72         & 50.50        & 90.00 \\
LbMAS (Qwen)       & 84.12 & 94.11         & 47.47        & 87.60 \\
\textit{- average} & 82.67 & 92.91         & 48.98        & 88.80 \\ \midrule
LbMAS              & 85.35 & 93.43         & 54.04        & 90.00 \\ \bottomrule
\end{tabular}
\label{tab:totalsingle}
\end{table*}

\begin{figure*}[htbp]
    \centering
    \includegraphics[width=1\linewidth]{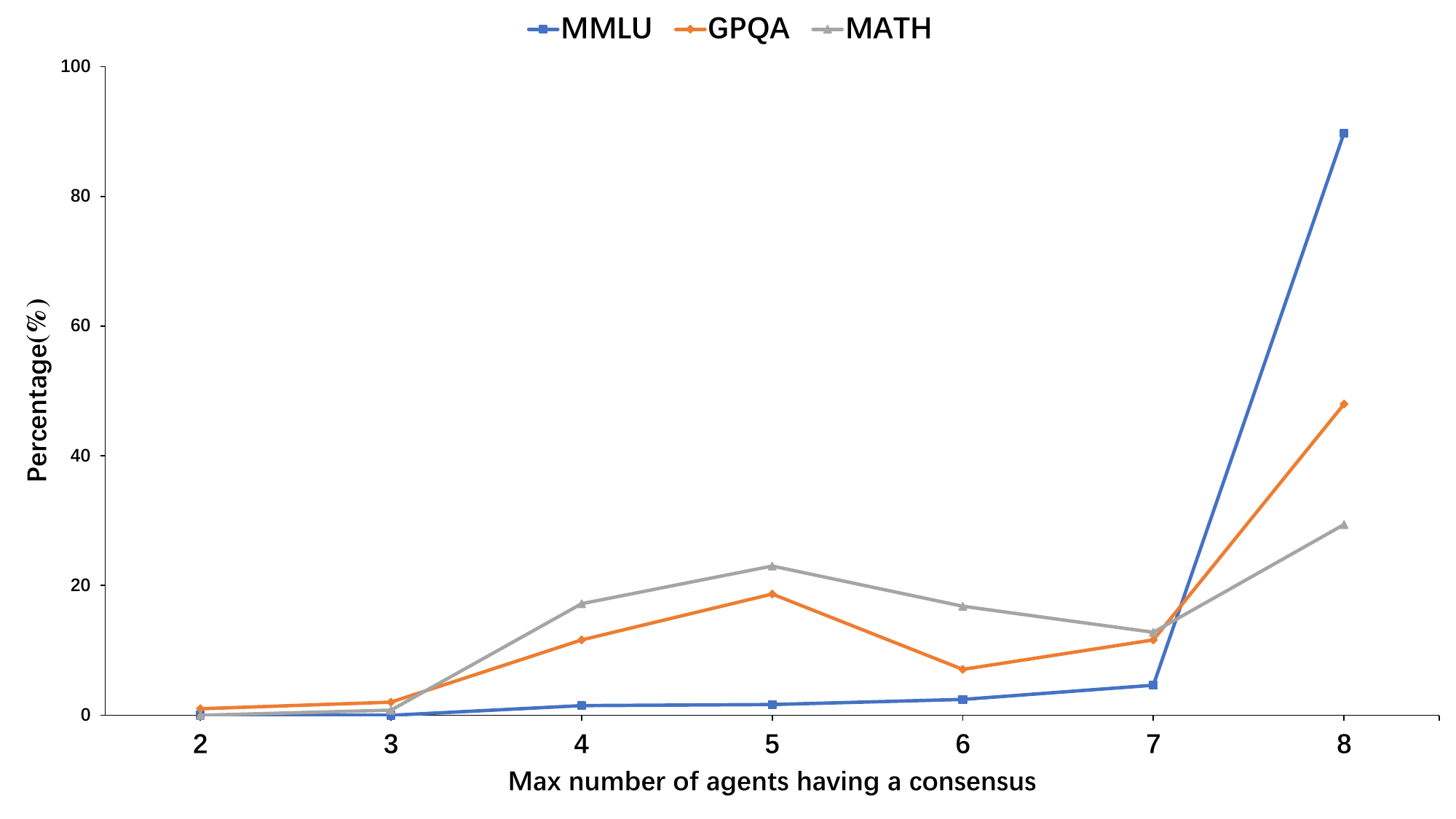}
    \caption{Max number of agents having a consensus in three datasets.}
    \label{fig:consensus}
\end{figure*}

\end{document}